\documentclass[12pt,a4paper,dvips]{article}
\usepackage{a4p}
\usepackage{cite,mcite,epsfig,rotating}
\usepackage{graphicx}
\usepackage{physics }
\usepackage{l3_title,ifthen,Lep}
\journalname{Phys. Lett. B}
\date{June 28, 2001}
\preprint{2001-045}
\newcommand{\rar}{\rightarrow}

\def\mm{\mbox{$M \hskip 0.010in \! \! \! \! \! / _{ll}$}}

\def\Nl{\ifmmode \mathrm{N_\ell} \else
                $\mathrm{N_\ell}$   \fi}%
\def\Ne{\ifmmode \mathrm{N}_e \else
                $\mathrm{N}_e$   \fi}%
\def\Nm{\ifmmode \mathrm{N}_\mu \else
                $\mathrm{N}_\mu$   \fi}%
\def\Nt{\ifmmode \mathrm{N}_\tau \else
                $\mathrm{N}_\tau$   \fi}%
\def\Ul{\ifmmode U_{\ell N} \else
                $U_{\ell N}$   \fi}%
\def\Ue{\ifmmode U_{eN} \else
                $U_{eN}$   \fi}%
\def\MZ{m_{\mathrm{Z}}}
\def\MW{m_{\mathrm{W}}}
\def\MN{m_{\mathrm{N}}}
\def\Mresc{m_{\mathrm{\it resc}}}
\def\MVI{m_{\mathrm{\it vis}}}
\def\Evi{E_{\mathrm{\it vis}}}

\def\Ne{\ifmmode \mathrm{N}_\mathrm{e}\else
                $\mathrm{N}_e$\fi}%
\def\Nm{\ifmmode \mathrm{N}_\mu\else
                $\mathrm{N}_\mu$\fi}%
\def\Nt{\ifmmode \mathrm{N}_\tau\else
                $\mathrm{N}_\tau$\fi}%

\newlength{\capindent}
\newlength{\capwidth}
\newlength{\figwidth}
\newcommand{\icaption}[2][!*!,!]{\hspace*{\capindent}%
  \begin{minipage}{\capwidth}
    \ifthenelse{\equal{#1}{!*!,!}}%
      {\caption{#2}}%
      {\caption[#1]{#2}}
  \end{minipage}}
\begin{document}
\bibliographystyle{l3style}
\begin{titlepage}
 \title{ Search for Heavy Isosinglet Neutrino \\
in e$^+$e$^-$ Annihilation at LEP  }
\author{The L3 Collaboration}
%
%
\begin{abstract}

We report on a search for the first generation 
heavy neutrino that is  an isosinglet
under the standard $SU(2)_L$ gauge group.
The data collected with the L3 detector 
at center-of-mass energies between
130 \GeV\ and   208 \GeV\ are used.
The decay channel $\Ne \rightarrow \mathrm{e}  \mathrm{W}$ is investigated
and no
 evidence is found for a heavy neutrino, $\Ne$,
 in a mass range between 80 \GeV\ 
and 205 \GeV. Upper
limits on the mixing parameter between the heavy and light neutrino
are derived.
   
\end{abstract}

\submitted 

 \end{titlepage}

\section*{Introduction}

In the Standard Model of electroweak interactions~\cite{SMM},
neutrinos are the only fundamental fermions which do not have a right-handed 
component that transforms as an isosinglet under the  $SU(2)_L$
 gauge group. 
However, additional  heavy isosinglet neutrinos
occur in various models that attempt 
to unify the presently known interactions into 
a single gauge scheme, such as Grand Unified Theories or Superstring inspired
models \cite{Valle2}. Several  extended 
electroweak models, including  left-right symmetric and see-saw 
models \cite{seesaw} also predict the existence of such neutrinos.

 Heavy isosinglet neutrinos can couple to the W and Z bosons through their 
mixing  with the light neutrinos. 
Constraints on 
 isosinglet neutrino
mixing were set by several experiments \cite{otherINHL,lepINHL,L3INHL}.
 Heavy
 neutrinos were searched for in 
leptonic decays of mesons and in neutrino
beam experiments~\cite{otherINHL}, resulting in stringent upper limits on 
the square of
their mixing amplitude to ordinary neutrinos,
 $|U_\ell|^2$, down to 10$^{-7}$ in the  mass 
region below 3~\GeV.  LEP experiments~\cite{lepINHL}  
 set
limits on $|U_\ell|^2$ of the order of 
$10^{-3}$~to~$10^{-5}$ for the neutrino mass range from 
3 \GeV\ up to 80 \GeV, and the 
 L3 experiment  derived the first limits on
$|U_\ell|^2$ for  neutrino masses  above the W mass~\cite{L3INHL}.

The data used in this analysis were collected with the L3 
detector~\cite{hlep04}
at LEP  at center-of-mass energies, $\sqrt{s}$, between
192 \GeV\ and   208 \GeV\
corresponding to an integrated  luminosity of  
450~pb$^{-1}$, out of which $\sim$115~pb$^{-1}$ were collected at 
$\sqrt{s}$ = 206.5~\GeV\ and $\sim$8~pb$^{-1}$ at $\sqrt{s}$~=~208~\GeV. 
Final results  also  include our 
 earlier data recorded at $\sqrt{s}~=~133-189$ \GeV~\cite{L3INHL}.

\section*{Production and decay}

This search is performed under the assumption that 
 one   heavy isosinglet neutrino \Nl is 
associated with each generation of light neutrinos with  the mixing amplitude
$U_\ell$. Neither the 
mixing between light neutrinos and higher isodoublet states
 nor the mixing among light neutrinos are 
considered~\cite{Gronau}.

In e$^+ $e$^-$ annihilation,  single production of
heavy neutrinos occurs via the mixing  between the heavy neutrino and
its  associated
isodoublet neutrino, as presented in  Figure~1:
\vskip 0.5cm
\centerline{e$^+ $e$^- \rightarrow \Nl  \nu_{\ell} $.}            
\vskip 0.5cm  
The 
corresponding heavy neutrino pair production cross section is suppressed
with respect  to the single production cross section by 
an additional $|U_\ell|^2$ factor, which is expected to be below
 0.1 for a heavy neutrino mass,  $\MN$,
 larger than
 80~\GeV~\cite{singlet_lim}.
The single production  proceeds through $s$-channel Z exchange for
all generations. In addition, the 
first generation heavy neutrinos, $\Ne$,  which couple to
electrons,  are also produced through  $t$-channel W exchange.
Figure~2 shows that the $t$-channel  contributions to the total
 production cross section are dominant and 
the production cross section for  $\Ne$  can be as
high as 0.7 pb.
The production cross section for~\Nm~and~\Nt~is below the sensitivity
of LEP and these heavy neutrinos are not considered 
in  the following.

Heavy isosinglet neutrinos decay via the neutral or charged weak currents:
\vskip 0.5cm
\[ \Ne
 \rightarrow \mathrm{Z}\nu_\mathrm{e} \mbox{~or~}
 \Ne \rightarrow \mathrm{eW}. \]
\vskip 0.5cm 

The decay into the
Z boson is suppressed by the limited phase space 
for heavy neutrinos with masses close to the  the W and Z masses.
 For  masses above 150~\GeV\ the branching ratios
reach the asymptotic values  
$Br(\Ne \rightarrow  \mathrm{eW}) = 2/3$ and
$Br(\Ne \rightarrow \mathrm{Z}\nu_\mathrm{e} ) = 1/3$~\cite{singlet_djo}.

\section*{Event simulation}

Using  the full differential cross section~\cite{singlet_buh},
 a dedicated Monte Carlo
generator is constructed to simulate the production and decay of 
the heavy isosinglet neutrinos.
 Subsequent hadronic fragmentation and decays are
simulated by the JETSET Monte Carlo program~\cite{hlep07}. The 
effects
of the finite width of the produced W and Z bosons as well as
initial and final state radiation are  taken into account.
This Monte Carlo program is used to generate several samples of signal
events with heavy neutrino masses ranging  from 80~\GeV\ 
up to the kinematic limit. 
For the simulation of background from Standard Model processes, the
following Monte Carlo programs are used:
KK2f~\cite{hlep06} ($\mbox{e}^+ \mbox{e}^-  \rar \mbox{q} \bar{\mbox{q}}
 (\gamma)$),
PYTHIA~\cite{hlep07} ($\mbox{e}^+ \mbox{e}^-  \rar 
\mbox{Z} \mbox{e}^+ \mbox{e}^-,~\mbox{Z} \mbox{Z}$),
KORALZ~\cite{hlep071} ($\mbox{e}^+ \mbox{e}^- \rar \tau^+ \tau^- (\gamma)$),
KORALW~\cite{hlep072} ($\mbox{e}^+ \mbox{e}^- \rar \mbox{W}^+ \mbox{W}^- $),
PHOJET~\cite{hlep08} ($\mbox{e}^+ \mbox{e}^- \rar \mbox{e}^+ \mbox{e}^- \mbox{q} \bar{\mbox{q}}$),
DIAG36~\cite{hlep081} ($\mbox{e}^+ \mbox{e}^- \rar \mbox{e}^+ \mbox{e}^- \tau^+ \tau^-$),
and EXCALIBUR~\cite{hlep082} for other four-fermion final states.

The Monte Carlo events are simulated in the L3 detector
using the GEANT~\cite{GEANT} and GHEISHA~\cite{GEISHA} programs,
 which take into account the
effects of energy loss, multiple scattering and showering in the
materials. Time dependent detector inefficiencies, as monitored
during the data taking, are also reproduced.

\section*{Event signatures and selection}

The present analysis concentrates
on  the decay channel $ \Ne \rightarrow\mathrm{eW}$
with $\mathrm{W} \rightarrow \mathrm{jets}$. The 
signature of these events
is one isolated electron plus hadronic jets. 
Since there is 
only one neutrino in the final state, it is possible to
 reconstruct 
 the invariant mass of the heavy neutrino, that
will manifest itself 
as  a peak in the invariant mass distribution.
Moreover, this decay channel has the largest branching ratio varying
between 68\% and 45\% depending on the heavy neutrino mass.
The dominant backgrounds come from  W$^+$W$^-$~production  
with one hadronic and one leptonic W decay (92\%),
$\mbox{q} \bar{\mbox{q}} (\gamma)$ (5\%) and ZZ production (2\%). 

The electron identification and jet reconstruction procedures follow
the criteria described  
in Reference~\citen{L3INHL}.
The event selection requires at 
least two hadronic jets plus one isolated electron.
The visible energy must exceed  70 \GeV\ and the number of
reconstructed tracks must be greater than 6. 
The polar angle $\theta$ of the missing momentum has to be in the range
$25^{\circ}<\theta<155^{\circ}$. 
The visible 
 mass  of the event, $\MVI$,  is reconstructed and, to improve the
resolution, it is rescaled as:
$$\Mresc=\MVI {\sqrt s\over p_\nu+\Evi}~,$$            
where $p_\nu$
is the missing momentum of the event, and $\Evi$ is the
visible energy. 
Figure~3 presents the distribution of the rescaled invariant
mass, $\Mresc$,    after the application of 
the  previous cuts.
 Good agreement is found between data and Monte Carlo expectation.
This spectrum is divided in two regions of $\Mresc$,  below and above 100~\GeV.
 In the first, ``region 1'', the heavy neutrino mass is close
to the W mass and  a significant fraction of W's produced in
$\Ne$ decays are off-shell. For $\Mresc$~$>$~100~\GeV, ``region 2'', 
 the W's are produced mostly on-shell.
In this case  a
 kinematic fit improves 
the resolution on the mass measurement, the determination
of jet energies and angles, and the missing momentum direction for both  
the signal 
and the W$^+$W$^-$ background.
 Four-momentum conservation and  the constraint that 
the invariant mass of the  hadronic jets is
equal to the W mass, are imposed in the fit.
In region 1  
we select 27 data events with    $23.6\pm0.6$ events 
expected from the 
Standard Model processes.
The corresponding numbers for region 2 are 794 and $776.2\pm3.5$.
Figure~4 displays the distribution of the invariant mass of the electron and
the  missing momentum, $m_{\mathrm{e} \nu}$, for events in region~2 after the 
application of the kinematic 
fit. A clear peak coming from 
the W$^+$W$^-$  background is observed at the W mass.

Finally, the  W$^+$W$^-$ background  is reduced by  requiring the
invariant mass of the electron and missing  momentum to be 
outside the W~mass region,
$m_{\mathrm{e} \nu}$~$<$~70~\GeV\ or $m_{\mathrm{e} \nu}$~$>$~90~\GeV, 
which  rejects 70\% of the  background events.
Figure~5 shows the invariant mass of the events accepted after this
 cut.
We observe a good agreement  between the data and expected Standard Model 
background: 233 data
events pass the selection with $226.5\pm1.8$  events
expected from the 
 background, out of which  
88\% are from  W$^+$W$^-$  production, 9\% from $\mbox{q}
 \bar{\mbox{q}} (\gamma)$
production and 3\% from ZZ production. 

\section*{Results}

As no signal is observed, 
the 95\% confidence level upper limits on the square of the
mixing amplitude, $|U_{\mathrm{e}}|^2$, are calculated from
the number of the data and background events~\cite{singlet_obr}. 
In region 1,  the total number of selected and expected events 
is used.
In region 2, the number of events
in data and  background for a given heavy neutrino 
mass $\MN$ is defined as the number of
events 
with a reconstructed mass in the interval  $\MN \pm 2\sigma$.
 The mass resolution $\sigma$ varies from~2~to~2.5~GeV over the 
investigated mass
 range.
The overall selection efficiency for heavy neutrino events 
varies smoothly from 20\% up to 45\% depending on
the values of $\MN$ and $\sqrt{s}$.
 The systematic uncertainty on the signal selection 
efficiency is mainly due to the uncertainty in
the simulation and reconstruction of the heavy neutrinos ($\sim$3\%),
the signal Monte Carlo statistics ($\sim$3\%),
and the energy calibration ($\sim$2\%). It
is estimated to be 5\% relative and  is
 taken into  account in the limit calculation by reducing the selection
 efficiency by 5\%.

Figure~6 shows the measured upper limits on the mixing amplitude
 $|U_{\mathrm{e}}|^2$ as a function of the heavy neutrino mass, along
with the expected limits as calculated from a large number of 
Monte Carlo experiments. These  results are obtained
using the whole data sample collected by L3 at LEP, and improve upon
and supersede our previously published results~\cite{L3INHL}.

\section*{Acknowledgements}

We wish to express our gratitude to the CERN accelerator divisions for the 
excellent performance of the LEP machine. We acknowledge with appreciation 
the effort of the engineers, technicians and support staff who have
participated in the construction and maintenance of this experiment.

%
%
\newpage
\section*{Author List}
\typeout{   }     
\typeout{Using author list for paper 237 -- ? }
\typeout{$Modified: Fri Jan 26 2001 by smele $}
\typeout{!!!!  This should only be used with document option a4p!!!!}
\typeout{   }
%
%
%
%
%
%

\newcount\tutecount  \tutecount=0
\def\tutenum#1{\global\advance\tutecount by 1 \xdef#1{\the\tutecount}}
\def\tute#1{$^{#1}$}
\tutenum\aachen            
\tutenum\nikhef            
\tutenum\mich              
\tutenum\lapp              
\tutenum\basel             
\tutenum\lsu               
\tutenum\beijing           
\tutenum\berlin            
\tutenum\bologna           
\tutenum\tata              
\tutenum\ne                
\tutenum\bucharest         
\tutenum\budapest          
\tutenum\mit               
\tutenum\panjab            
\tutenum\debrecen          
\tutenum\florence          
\tutenum\cern              
\tutenum\wl                
\tutenum\geneva            
\tutenum\hefei             
\tutenum\lausanne          
\tutenum\lyon              
\tutenum\madrid            
\tutenum\florida           
\tutenum\milan             
\tutenum\moscow            
\tutenum\naples            
\tutenum\cyprus            
\tutenum\nymegen           
\tutenum\caltech           
\tutenum\perugia           
\tutenum\peters            
\tutenum\cmu               
\tutenum\potenza           
\tutenum\prince            
\tutenum\riverside         
\tutenum\rome              
\tutenum\salerno           
\tutenum\ucsd              
\tutenum\sofia             
\tutenum\korea             
\tutenum\utrecht           
\tutenum\purdue            
\tutenum\psinst            
\tutenum\zeuthen           
\tutenum\eth               
\tutenum\hamburg           
\tutenum\taiwan            
\tutenum\tsinghua          

{
\parskip=0pt
\noindent
{\bf The L3 Collaboration:}
\ifx\selectfont\undefined
 \baselineskip=10.8pt
 \baselineskip\baselinestretch\baselineskip
 \normalbaselineskip\baselineskip
 \ixpt
\else
 \fontsize{9}{10.8pt}\selectfont
\fi
\medskip
\tolerance=10000
\hbadness=5000
\raggedright
\hsize=162truemm\hoffset=0mm
\def\r{\rlap,}
\noindent

P.Achard\r\tute\geneva\ 
O.Adriani\r\tute{\florence}\ 
M.Aguilar-Benitez\r\tute\madrid\ 
J.Alcaraz\r\tute{\madrid,\cern}\ 
G.Alemanni\r\tute\lausanne\
J.Allaby\r\tute\cern\
A.Aloisio\r\tute\naples\ 
M.G.Alviggi\r\tute\naples\
H.Anderhub\r\tute\eth\ 
V.P.Andreev\r\tute{\lsu,\peters}\
% ...
\newpage
\begin{list}{A}{\itemsep=0pt plus 0pt minus 0pt\parsep=0pt plus 0pt minus 0pt
                \topsep=0pt plus 0pt minus 0pt}
\item[\aachen]
 I. Physikalisches Institut, RWTH, D-52056 Aachen, FRG$^{\S}$\\
 III. Physikalisches Institut, RWTH, D-52056 Aachen, FRG$^{\S}$
\item[\nikhef] National Institute for High Energy Physics, NIKHEF, 
     and University of Amsterdam, NL-1009 DB Amsterdam, The Netherlands
\item[\mich] University of Michigan, Ann Arbor, MI 48109, USA
\item[\lapp] Laboratoire d'Annecy-le-Vieux de Physique des Particules, 
     LAPP,IN2P3-CNRS, BP 110, F-74941 Annecy-le-Vieux CEDEX, France
\item[\basel] Institute of Physics, University of Basel, CH-4056 Basel,
     Switzerland
\item[\lsu] Louisiana State University, Baton Rouge, LA 70803, USA
\item[\beijing] Institute of High Energy Physics, IHEP, 
  100039 Beijing, China$^{\triangle}$ 
\item[\berlin] Humboldt University, D-10099 Berlin, FRG$^{\S}$
\item[\bologna] University of Bologna and INFN-Sezione di Bologna, 
     I-40126 Bologna, Italy
\item[\tata] Tata Institute of Fundamental Research, Mumbai (Bombay) 400 005, India
\item[\ne] Northeastern University, Boston, MA 02115, USA
\item[\bucharest] Institute of Atomic Physics and University of Bucharest,
     R-76900 Bucharest, Romania
\item[\budapest] Central Research Institute for Physics of the 
     Hungarian Academy of Sciences, H-1525 Budapest 114, Hungary$^{\ddag}$
\item[\mit] Massachusetts Institute of Technology, Cambridge, MA 02139, USA
\item[\panjab] Panjab University, Chandigarh 160 014, India.
\item[\debrecen] KLTE-ATOMKI, H-4010 Debrecen, Hungary$^\P$
\item[\florence] INFN Sezione di Firenze and University of Florence, 
     I-50125 Florence, Italy
\item[\cern] European Laboratory for Particle Physics, CERN, 
     CH-1211 Geneva 23, Switzerland
\item[\wl] World Laboratory, FBLJA  Project, CH-1211 Geneva 23, Switzerland
\item[\geneva] University of Geneva, CH-1211 Geneva 4, Switzerland
\item[\hefei] Chinese University of Science and Technology, USTC,
      Hefei, Anhui 230 029, China$^{\triangle}$
\item[\lausanne] University of Lausanne, CH-1015 Lausanne, Switzerland
\item[\lyon] Institut de Physique Nucl\'eaire de Lyon, 
     IN2P3-CNRS,Universit\'e Claude Bernard, 
     F-69622 Villeurbanne, France
\item[\madrid] Centro de Investigaciones Energ{\'e}ticas, 
     Medioambientales y Tecnolog{\'\i}cas, CIEMAT, E-28040 Madrid,
     Spain${\flat}$ 
\item[\florida] Florida Institute of Technology, Melbourne, FL 32901, USA
\item[\milan] INFN-Sezione di Milano, I-20133 Milan, Italy
\item[\moscow] Institute of Theoretical and Experimental Physics, ITEP, 
     Moscow, Russia
\item[\naples] INFN-Sezione di Napoli and University of Naples, 
     I-80125 Naples, Italy
\item[\cyprus] Department of Physics, University of Cyprus,
     Nicosia, Cyprus
\item[\nymegen] University of Nijmegen and NIKHEF, 
     NL-6525 ED Nijmegen, The Netherlands
\item[\caltech] California Institute of Technology, Pasadena, CA 91125, USA
\item[\perugia] INFN-Sezione di Perugia and Universit\`a Degli 
     Studi di Perugia, I-06100 Perugia, Italy   
\item[\peters] Nuclear Physics Institute, St. Petersburg, Russia
\item[\cmu] Carnegie Mellon University, Pittsburgh, PA 15213, USA
\item[\potenza] INFN-Sezione di Napoli and University of Potenza, 
     I-85100 Potenza, Italy
\item[\prince] Princeton University, Princeton, NJ 08544, USA
\item[\riverside] University of Californa, Riverside, CA 92521, USA
\item[\rome] INFN-Sezione di Roma and University of Rome, ``La Sapienza",
     I-00185 Rome, Italy
\item[\salerno] University and INFN, Salerno, I-84100 Salerno, Italy
\item[\ucsd] University of California, San Diego, CA 92093, USA
\item[\sofia] Bulgarian Academy of Sciences, Central Lab.~of 
     Mechatronics and Instrumentation, BU-1113 Sofia, Bulgaria
\item[\korea]  The Center for High Energy Physics, 
     Kyungpook National University, 702-701 Taegu, Republic of Korea
\item[\utrecht] Utrecht University and NIKHEF, NL-3584 CB Utrecht, 
     The Netherlands
\item[\purdue] Purdue University, West Lafayette, IN 47907, USA
\item[\psinst] Paul Scherrer Institut, PSI, CH-5232 Villigen, Switzerland
\item[\zeuthen] DESY, D-15738 Zeuthen, 
     FRG
\item[\eth] Eidgen\"ossische Technische Hochschule, ETH Z\"urich,
     CH-8093 Z\"urich, Switzerland
\item[\hamburg] University of Hamburg, D-22761 Hamburg, FRG
\item[\taiwan] National Central University, Chung-Li, Taiwan, China
\item[\tsinghua] Department of Physics, National Tsing Hua University,
      Taiwan, China
\item[\S]  Supported by the German Bundesministerium 
        f\"ur Bildung, Wissenschaft, Forschung und Technologie
\item[\ddag] Supported by the Hungarian OTKA fund under contract
numbers T019181, F023259 and T024011.
\item[\P] Also supported by the Hungarian OTKA fund under contract
  number T026178.
\item[$\flat$] Supported also by the Comisi\'on Interministerial de Ciencia y 
        Tecnolog{\'\i}a.
\item[$\sharp$] Also supported by CONICET and Universidad Nacional de La Plata,
        CC 67, 1900 La Plata, Argentina.
\item[$\triangle$] Supported by the National Natural Science
  Foundation of China.
\end{list}
}
\vfill


\newpage


\newpage


\clearpage
\vspace*{5cm}

\input{feynman}\bigphotons      
%
%
\begin{figure}[htbp]
\begin{center}
\mbox{\normalsize
\begin{picture}(15000,35000)(-15000,-42000)
\THICKLINES    
\drawline\photon[\E\REG](-15000,0)[7]
\global\advance\pmidx by -200
\global\advance\pmidy by  1000
\put(\pmidx,\pmidy){{\Large Z}}

\drawline\fermion[\NE\REG](\photonbackx,\photonbacky)[5500]
\global\advance\photonbackx by 50
\drawline\fermion[\NE\REG](\photonbackx,\photonbacky)[5500]
\global\advance\photonbackx by 50
\drawline\fermion[\NE\REG](\photonbackx,\photonbacky)[5500]
\global\advance\photonbackx by 50
\drawline\fermion[\NE\REG](\photonbackx,\photonbacky)[5500]
\global\advance\photonbackx by -200
\global\advance\pmidx by +2400
\global\advance\pmidy by +2000
\put(\pmidx,\pmidy){{\Large N$_\ell$}}
\global\advance\pmidx by -3250

\global\advance\pmidy by -4500
\put(\pmidx,\pmidy){{\large ${U_{\ell}}$}}

\put(\photonbackx,\photonbacky){\circle*{750}}
\drawline\fermion[\SE\REG](\photonbackx,\photonbacky)[5500]
\global\advance\photonbackx by 500
\global\advance\photonbacky by -1500
\global\advance\pmidx by 2200 
\global\advance\pmidy by -3200
\put(\pmidx,\pmidy){{\Large $\nu_\ell$}}
\drawline\photon[\E\REG](\photonfrontx,\photonfronty)[6]
\global\advance\pmidx by +1600
\global\advance\pmidy by +1000
\global\advance\pmidx by -1000
\drawline\fermion[\SW\REG](-15000,0)[5500]
\global\advance\pmidx by -4000
\global\advance\pmidy by -3200
\put(\pmidx,\pmidy){{\Large e$^-$}}
\drawline\fermion[\NW\REG](-15000,0)[5500]
\global\advance\pmidx by -3800
\global\advance\pmidy by +2000
\put(\pmidx,\pmidy){{\Large e$^+$}}
\global\advance\pmidx by -4800
\global\advance\pmidy by +2000
\put(\pmidx,\pmidy){\Large a)}
\global\advance\pmidy by -16000
\put(\pmidx,\pmidy){\Large b)}

\drawline\fermion[\NW\REG](-11000,-18000)[8000]
\global\advance\pmidx by -4000
\global\advance\pmidy by +3200
\put(\pmidx,\pmidy){{\Large e$^+$}}

\drawline\fermion[\NE\REG](-11100,-18000)[8000]
\drawline\fermion[\NE\REG](-11050,-18000)[8000]
\drawline\fermion[\NE\REG](-10950,-18000)[8000]
\drawline\fermion[\NE\REG](-11000,-18000)[8000]

\global\advance\pmidx by 3100
\global\advance\pmidy by 3200
\put(\pmidx,\pmidy){{\Large $\mathrm{N}_{\mathrm{e}}$}}
\global\advance\pmidx by -6650
\global\advance\pmidy by -4500

\put(\pmidx,\pmidy){{\large ${U_{\mathrm{e}}}$}}
\drawline\photon[\S\REG](-11000,-18000)[4]
\put(\photonfrontx,\photonfronty){\circle*{750}}
\drawline\photon[\S\REG](-11050,-18000)[4]

\global\advance\pmidx by -2200

\global\advance\pmidy by -2850
\put(\pmidx,\pmidy){{\Large W}}

\drawline\photon[\S\REG](\photonbackx,\photonbacky)[4]
\global\advance\photonfrontx by 50
\drawline\photon[\S\REG](\photonfrontx,\photonfronty)[4]

\global\advance\pmidx by -3200
\global\advance\pmidy by -850

\drawline\fermion[\SW\REG](\photonbackx,\photonbacky)[8000]
\global\advance\pmidx by -4000
\global\advance\pmidy by -4250
\put(\pmidx,\pmidy){{\Large e$^-$}}
\drawline\fermion[\SE\REG](\photonbackx,\photonbacky)[8000]
\global\advance\pmidx by 3100
\global\advance\pmidy by -4050
\put(\pmidx,\pmidy){{\Large ${\nu_{\mathrm{e}}}$}}
\global\advance\photonbackx by 500
\global\advance\photonbackx by 500
\global\advance\photonbacky by -500

\end{picture}
     }
\caption
{
Feynman diagrams showing the production of isosinglet neutrinos
via 
a) $s$-channel and b) $t$-channel. Here  $\ell$ denotes e,
$\mu$ or $\tau$ for the $s$-channel production.
}
\label{fey}
\end{center}
\end{figure}
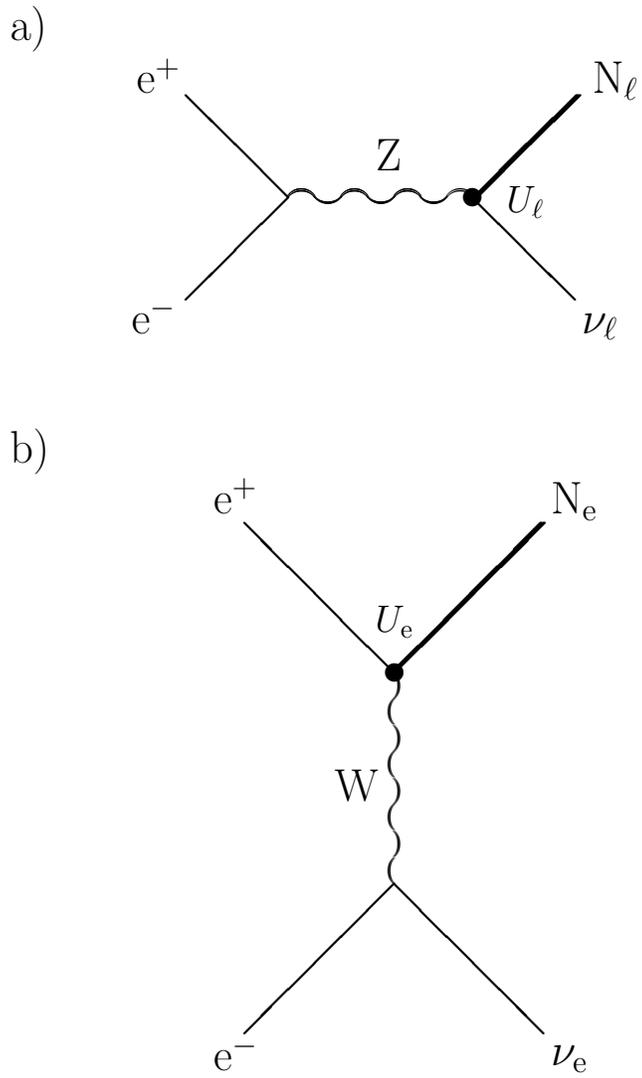

\begin{figure}[p]
\begin{center}
\mbox{\epsfysize=16cm\epsffile{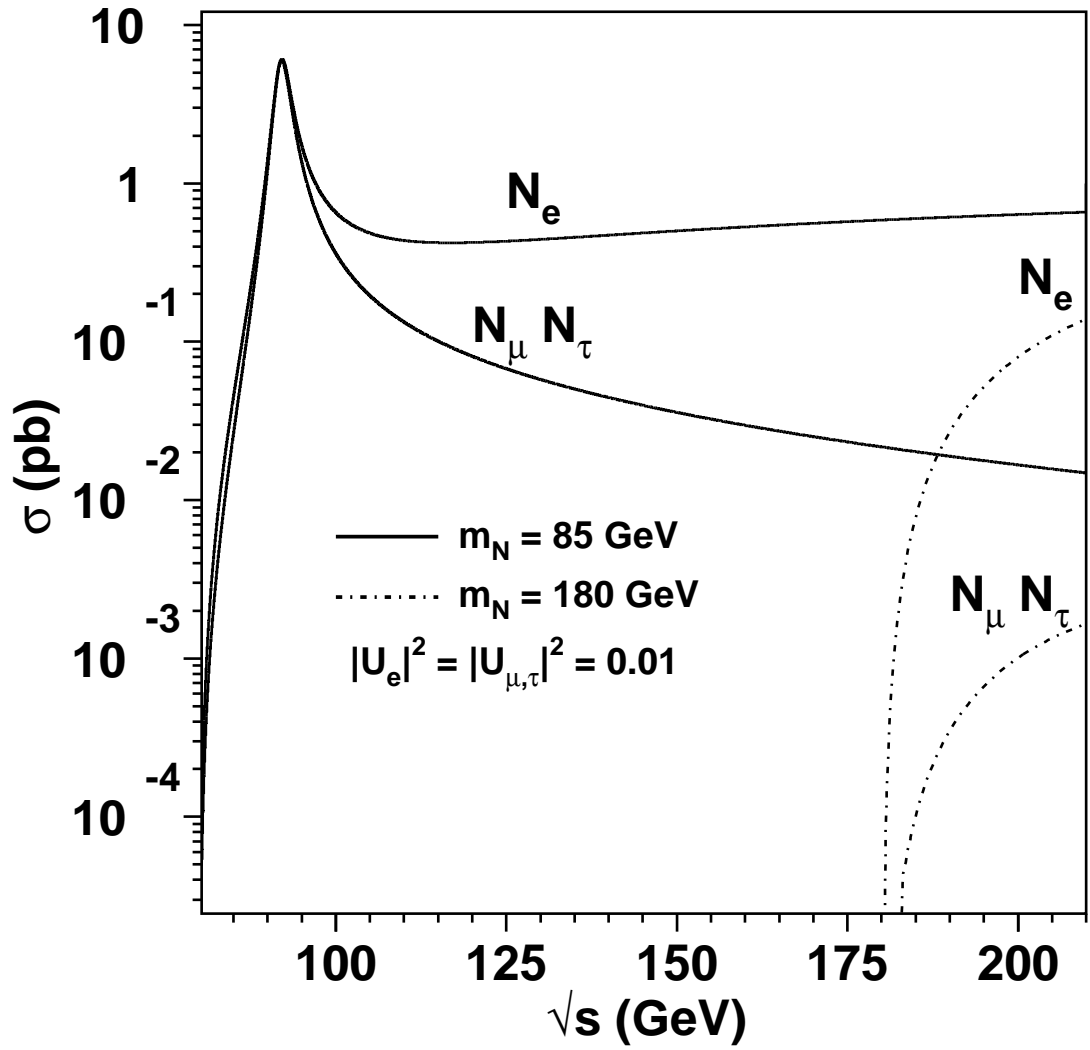}}
\end{center}
\caption
{
Total cross section for single production of heavy isosinglet
neutrinos, e$^+$e$^- \rightarrow \Nl \nu_{\ell}$, as a function
of the center-of-mass energy~[11].
}
\label {fig2_new_new}
\end{figure}

\begin{figure}[p]
\begin{center}
\mbox{\epsfysize=16cm\epsffile{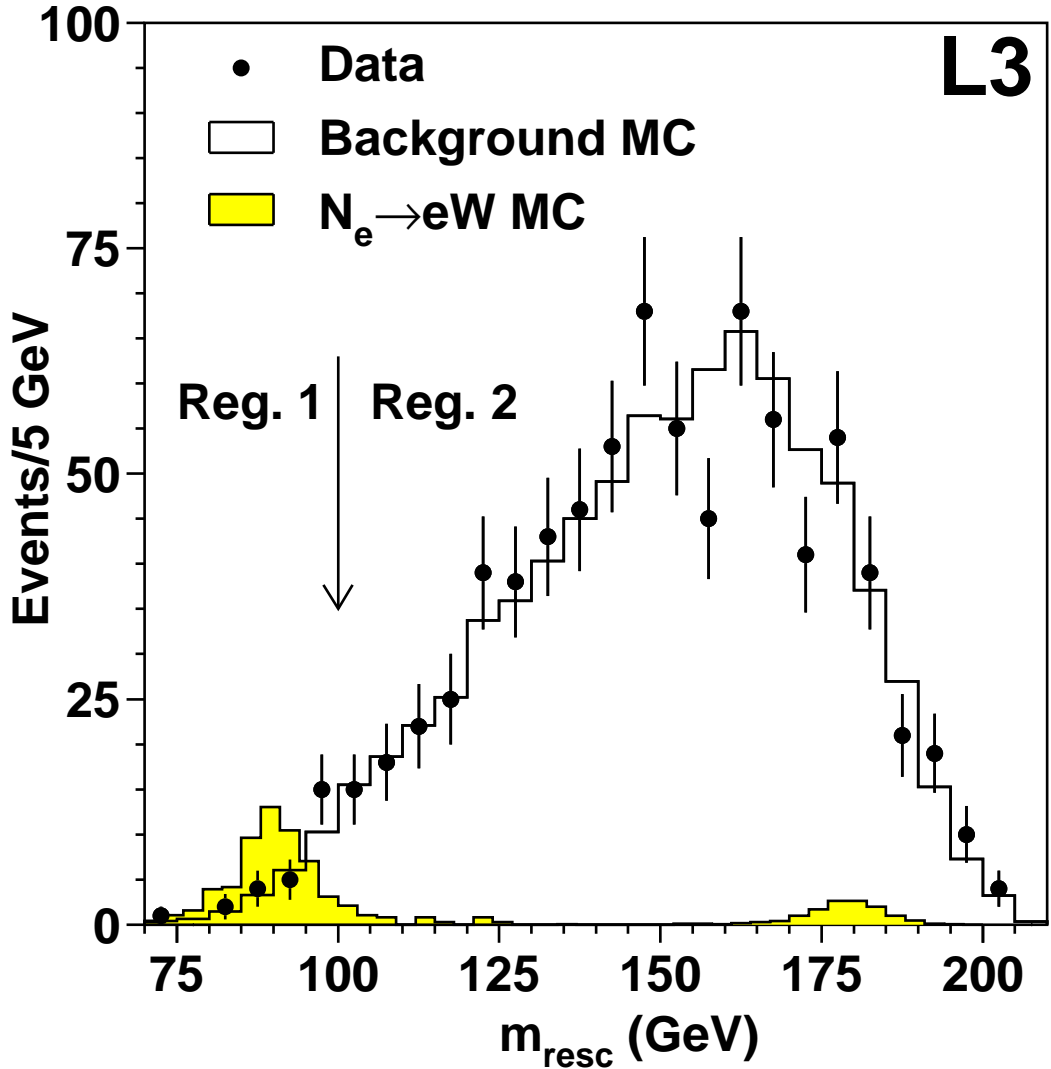}}
\end{center}
\caption
{
Distribution of the rescaled invariant mass, $\Mresc$, of the event. 
The points are the data, 
collected at $\sqrt{s}$ = 192$-$208 \GeV,
and the solid histogram 
is the background Monte Carlo. The shaded histogram is the predicted
 e$^+ $e$^- \rightarrow \nu \mathrm{N}_\mathrm{e}$ signal for the heavy
neutrino  masses  of 90~\GeV\ and 180~\GeV\ with the  mixing amplitude
 $|U_{\mathrm{e}}|$~=~0.1.
Both histograms are normalised to the same luminosity as the data.
The split of the spectrum into ``region~1'' and ``region~2'' is 
described in the  text.
}
\label {ainvs_all.eps}
\end{figure}

\begin{figure}[p]
\begin{center}
\mbox{\epsfysize=16cm\epsffile{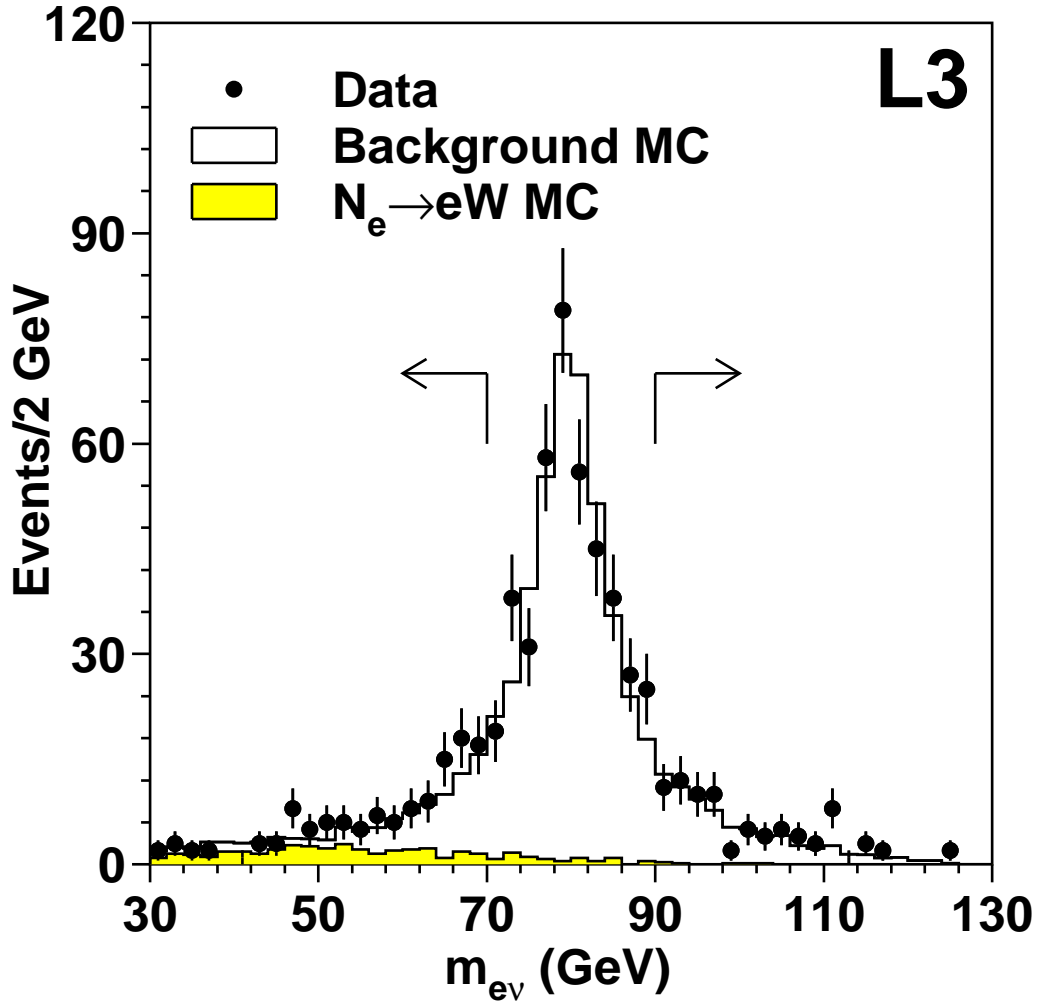}}
\end{center}
\caption
{
The invariant mass, $m_{\mathrm{e} \nu}$, of the isolated electron and 
missing momentum.
The points are the data, collected at $\sqrt{s}$ = 192$-$208 \GeV, 
and the solid histogram
is the background Monte Carlo. The shaded histogram is the predicted
 e$^+ $e$^- \rightarrow \nu \mathrm{N}_\mathrm{e}$ signal for a 150 \GeV\ heavy
neutrino  with the mixing amplitude
 $|U_{\mathrm{e}}|$~=~0.1. 
For better visibility, the normalization for the signal  is scaled
 by a factor of 2.
The arrows indicate the accepted  range of $m_{\mathrm{e} \nu}$
 outside the  70$-$90~\GeV\ window.
}
\label {ainwl}
\end{figure}

\begin{figure}[p]
\begin{center}
\mbox{\epsfysize=16cm\epsffile{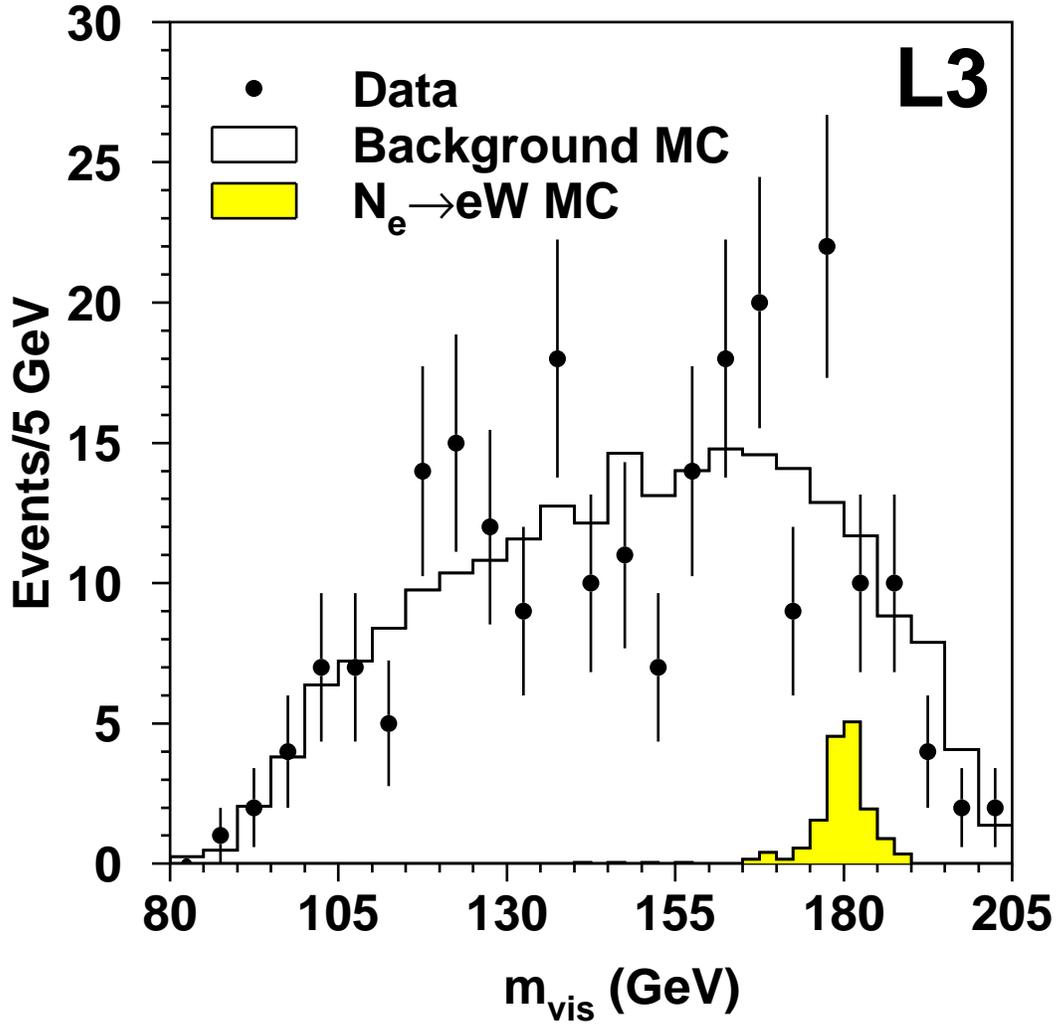}}
\end{center}
\caption
{
Distribution of the visible invariant mass of the event, $\MVI$,
after the kinematic fit. The points are
the data,
collected at $\sqrt{s}$ = 192$-$208 \GeV, 
and the solid histogram is the background Monte Carlo.
The shaded histogram is the predicted
 e$^+ $e$^- \rightarrow \nu \mathrm{N}_\mathrm{e}$ signal for a 180~\GeV\ heavy
neutrino with the mixing amplitude
 $|U_{\mathrm{e}}|$~=~0.1. For better visibility, 
the normalization for the signal  is scaled by a factor of 2.
}
\label {antot_all}
\end{figure}

\begin{figure}[p]
\begin{center}
\mbox{\epsfysize=16cm\epsffile{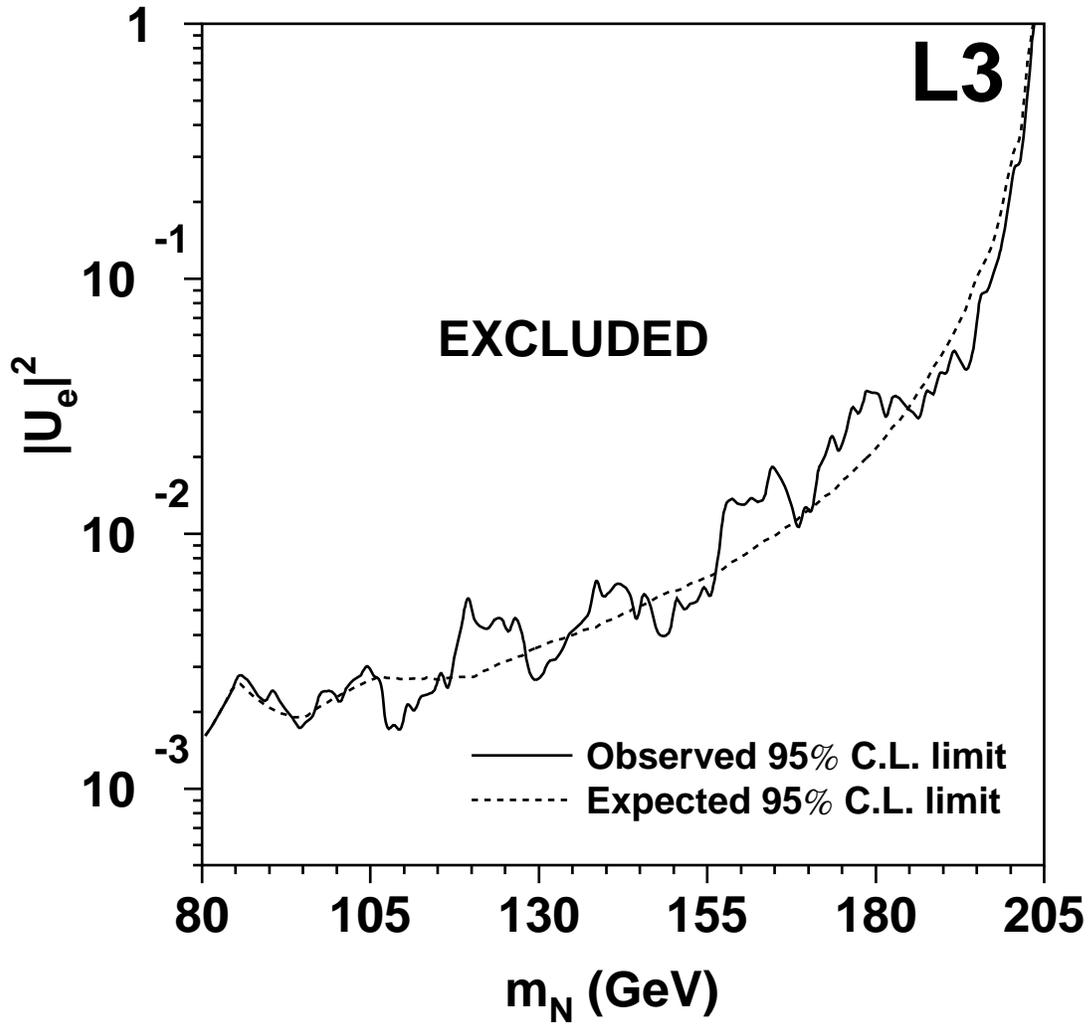}}
\end{center}
\caption
{
Observed and expected upper limits at the 95\% confidence level
 on the mixing amplitude $|U_{\mathrm{e}}|^2$
  as a function
of the  heavy isosinglet neutrino mass $\MN$.
}
\label {limit_maj}
\end{figure}

\end{document}